\documentstyle[aps,preprint,epsf]{revtex}
\begin{document} 
\preprint{WIS -- 98/17/June -- DPP}
\draft
\title{Density-matrix approach to coherent transport 
        and the measurement problem\\
\vskip1cm
{\normalsize Talk given
     at the V. Workshop on Nonequilibrium Physics at 
Short  - Time Scales, Rostock, April 27-30, 1998}}
\author {S.A. Gurvitz}
\address {Department of Particle Physics, Weizmann Institute of
       Science, Rehovot 76100, Israel}
\maketitle
\begin{abstract}
Bloch-type equations for description of coherent transport in
mesoscopic systems are applied for a study of the continuous measurement 
process. Both the detector and the measured system are 
described quantum mechanically. It is shown that 
the Schr\"odinger evolution of the entire 
system cannot be accommodated with the measurement collapse. 
The latter leads to quantum jumps which can be experimentally 
detected.   
\end{abstract}
\vskip 1cm
In a recent elegant experiment Buks {\em et al.}\cite{buks} 
realized a nondistructive continuous monitoring of a quantum system 
in the linear superposition by using
the ballistic point-contact as a detector\cite{buks}. 
Experiments of 
this type are very important for understanding of the 
measurement process, in particular since the quantum-mechanical behavior 
of the detector can be traced out. The latter allows us to 
investigate of whether the measurement collapse generates 
experimentally observed effects, which are not described 
by the Schr\"odinger equation applied to the entire system.    
As an example we consider continuous monitoring of a {\em single} 
electron inside the coupled-dot (Figs.~1)\cite{gur1}.
The point-contact (detector), 
shown as a barrier, is placed near one of the dots. 
The barrier is connected with two reservoirs at the chemical potentials 
$\mu_L$ and $\mu_R$ respectively. 
Since $\mu_L > \mu_R$, the current $I=eD$ 
flows through the point-contact, where $D=T(\mu_L-\mu_R)/(2\pi )$ and 
$T=(2\pi)^2\Omega^2\rho_L\rho_L$ is the transmission coefficient. Here 
$\Omega$ is the coupling between the left and the right reservoirs and 
$\rho_{L,R}$ is the corresponding density of the states.   
The penetrability of the point-contact (the barrier height)
is modulated by the electron, 
oscillating inside the double-dot.  
When the electron occupies the left dot, the transmission coefficient is 
$T_1$. However, when the 
right dot is occupied, the transmission coefficient 
$T_2\ll T_1$ due to the electrostatic repulsion generated by the electron. 
As a result, the current $I_2\ll I_1$. (We assume  
that $T_2=0$, so that the point contact is blocked whenever the 
right dot is occupied). 
Since the difference $\Delta I=I_1-I_2$
is macroscopically large, one can determine which of the dots 
is occupied by observing the point-contact current. 
Yet, the entire system can be treated quantum-mechanically. 
\vskip1cm
\begin{minipage}{13cm}
\begin{center}
\leavevmode
\epsfxsize=13cm
\epsffile{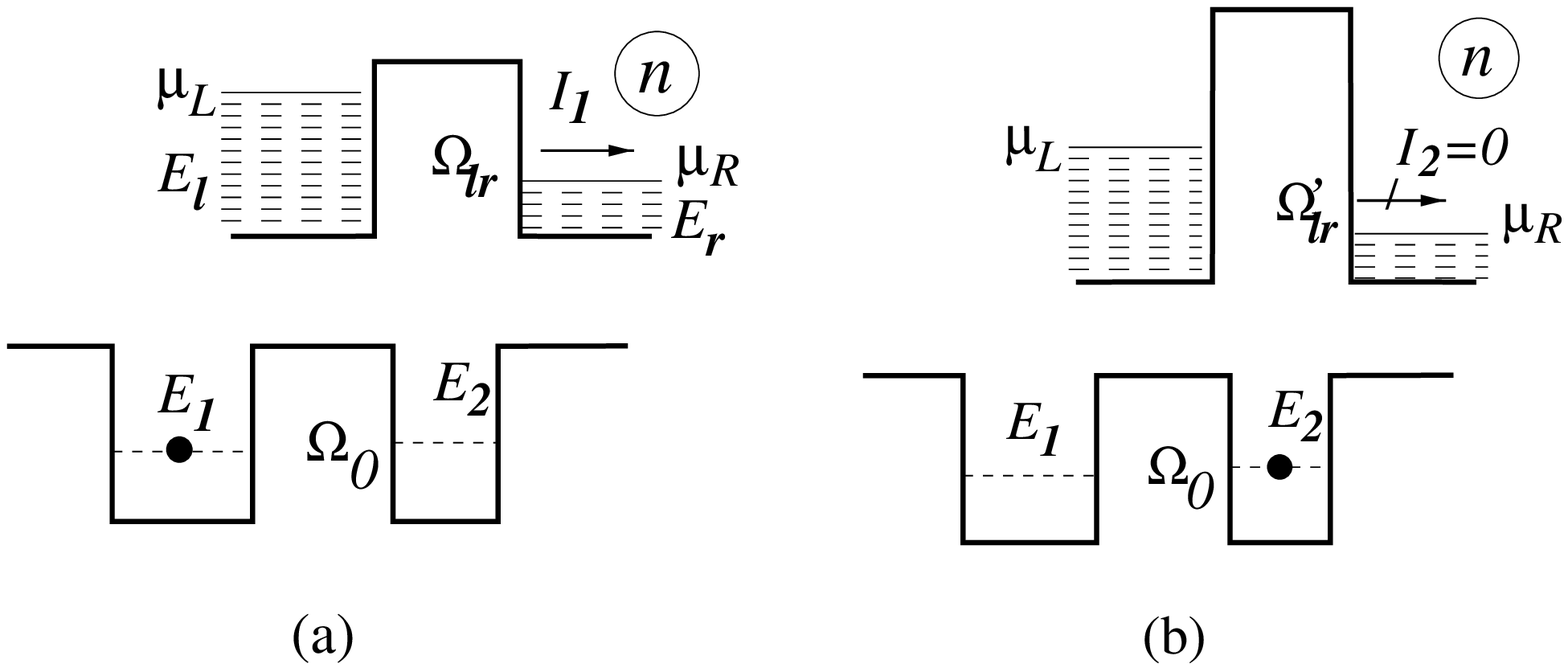}
\end{center}
{\begin{small}
Fig.~1. The point-contact detector near the double-dot.
$\Omega_{lr}$ is the coupling between the level $E_l$ and 
$E_r$ in the left and the right 
reservoirs. $\Omega_0$ is the coupling between the quantum dots.
The index $n$ denotes the number of electrons penetrating 
to the right reservoir (collector) at time $t$.
\end{small}} 
\end{minipage} \\ \\  \\
It is described by the Hamiltonian: 
${\cal H}={\cal H}_{PC}+{\cal H}_{DD}+{\cal H}_{int}$, where 
\begin{eqnarray}
{\cal H}_{PC}&=&\sum_l E_la_l^\dagger a_l+\sum_r E_ra_r^\dagger a_r
+\sum_{l,r}\Omega_{lr}(a_l^\dagger a_r +a_r^\dagger a_l )\, , 
\label{a2a}\\
{\cal H}_{DD} &=& E_1 c_1^{\dagger}c_{1}+E_2 c_2^{\dagger}c_{2}+
              \Omega_0 (c_2^{\dagger}c_{1}+ c_1^{\dagger}c_{2})\, ,
\label{a2b}\\
{\cal H}_{int}&=&\sum_{l,r}\delta\Omega_{lr}c_2^{\dagger} 
c_2(a^{\dagger}_la_r +a^{\dagger}_ra_l)\, .
\label{a2c}
\end{eqnarray}
Here ${\cal H}_{PC}$, ${\cal H}_{DD}$ and ${\cal H}_{int}$ are the 
Hamiltonians describing the point-contact, double-dot and their 
mutual interaction, respectively. The latter affects the coupling 
between the reservoirs. It becomes 
$\Omega'_{lr}=\Omega_{lr}+\delta\Omega_{lr}$ 
whenever the second dot is occupied. 
(In our case $\delta\Omega_{lr}=-\Omega_{lr}$).

The time-development of the entire system is described 
by the many-body Schr\"odinger equation 
$i\dot\rho (t) =[{\cal H},\rho (t)]$, where $\rho (t)$ is the total 
density-matrix. It was shown\cite{gur1} that the continuum reservoirs states
can be integrated out in the density-matrix $\rho$. Then the equation 
of motion becomes a system of coupled Bloch-type equations for the 
reduced density-matrix $\sigma (t)$: 
\begin{eqnarray}
\dot\sigma_{11}^{(n)} & = & -D_1\sigma_{11}^{(n)}+D_1\sigma_{11}^{(n-1)}
+i\Omega_0 (\sigma_{12}^{(n)}-\sigma_{21}^{(n)})\;, 
\label{c3a}\\
\dot\sigma_{22}^{(n)} & = & 
-i\Omega_0 (\sigma_{12}^{(n)}-\sigma_{21}^{(n)})\;,
\label{c3b}\\
\dot\sigma_{12}^{(n)} & = & i(E_2-E_1)\sigma_{12}^{(n)}+
i\Omega_0(\sigma_{11}^{(n)}-\sigma_{22}^{(n)})
-(1/2)D_1\sigma_{12}^{(n)}.
\label{c3c}
\end{eqnarray} 
Here  $\sigma^{(n)}_{11}(t)$, $\sigma^{(n)}_{22}(t)$ are 
the probabilities of finding the left or the right dot occupied,
with $n$ electrons in the collector.   
$\sigma^{(n)}_{12}(t)$ is the corresponding off-diagonal density-matrix 
element. 
Eqs.~(\ref{c3a}-\ref{c3c}) allows 
detailed microscopic study of the measurement process. 
For instance, the influence of the
detector on the measured system is determined by tracing out 
the detector states $n$. One finds
\begin{eqnarray}
\dot{\sigma}_{11}& = &i\Omega_0(\sigma_{12}-\sigma_{21})\;, 
\label{a6a}\\
\dot{\sigma}_{12}& = & i(E_2-E_1)\sigma_{12}+i\Omega_0(2\sigma_{11}-1)
-(1/2)D_1\sigma_{12}.
\label{a6c}
\end{eqnarray}   
where $\sigma_{ij}=\sum_n\sigma^{(n)}_{ij}$, 
and $\sigma_{22}=1-\sigma_{11}$.
As expected, the electron oscillations inside the double-dot are 
damped via the last (decoherence) 
term in Eq.~(\ref{a6c}), generated by the detector. Then 
the reduced electron density-matrix $\sigma_{ij}(t)$ 
becomes the statistical mixture for $t\to\infty$: 
\begin{equation} 
\sigma (t)=\left (\begin{array}{cc}
\sigma_{11}(t)&\sigma_{12}(t)\\
\sigma_{21}(t)&\sigma_{22}(t)\end{array}\right )
\to\left (\begin{array}{cc}
1/2&0\\0&1/2
\end{array}\right ) \;\;\; {\mbox{for}}\;\;\;  t\gg t_0\, ,
\label{a8}
\end{equation}  
Yet, the relaxation time $t_0$ {\em increases} 
with the dephasing rate $D_1$.
It follows from Eqs.~(\ref{a6a})-(\ref{a6c}) 
that $t_0\simeq D_1/8\Omega_0^2$, so that 
the continuous measurement slows down 
the transition rate between different states of the observed system. 
This result looks as a manifestation of the 
measurement collapse (Zeno effect). Yet, it was obtained from 
the continuous Schr\"odinger evolution of the entire system without 
any explicit relation to the collapse (cf. \cite{fre}).

Nevertheless, the problem arises with evaluation of 
the detector current.
Consider for instance, the case when 
the electron density-matrix becomes the statistical mixture, 
Eq.~(\ref{a8}). On the first sight one can expect 
that the detector current 
would display the average value, $I_1/2$. On the other hand,  
the mixture means that the electron actually 
occupies one of the dots, and therefore the detector should 
show either the current $I_1$ or $0$, but not the average. 
In fact, the behavior of the detector 
current cannot be determined from the reduced electron 
density-matrix, Eq.~(\ref{a8}). We have to study the total density-matrix 
$\sigma_{ij}^{(n)}$,  
Eqs.~(\ref{c3a})-(\ref{c3c}), which provide quantum-mechanical description 
of the entire system, including the detector\cite{gurv2}. 
First consider the case of $\Omega_0=0$, so that the electron 
is permanently localized in the left well. Solving Eq.~(\ref{c3a}) by 
using the Fourier transform\cite{sasha} we find 
\begin{equation} 
\sigma^{(n)}_{11}(t)=\bar\sigma^{(n)}(t)\simeq 
1/(2\pi D_1t)^{1/2}
\exp\left [-(D_1t-n)^2/2D_1t\right ]
\label{a2}
\end{equation}
where $\bar\sigma^{(n)}(t)$ is the probability of finding
$n$ electrons in the collector. The number $n$ increases with the 
rate $D_1$ that determines the detector current. 
Now we are going to general case of $\Omega_0\not =0$.
Consider again $D_1\gg\Omega_0$, 
so that the electron, initially localized in one of the well stays 
there for a long time ($t_0$). Solving numerically
Eqs.~(\ref{c3a})-(\ref{c3c}) with the initial conditions
$\sigma_{11}^{(0)}(0)=1$, $\sigma_{12}^{(0)}(0)=\sigma_{22}^{(0)}(0)=0$,
we find that $\sigma_{11}^{(n)}(t)\simeq\bar\sigma^{(n)}(t)$
for $t{\lower2pt\hbox{$\stackrel{<}{\sim}$}}t_0$ 
and $\sigma_{12}^{(n)}(t)$ and $\sigma_{22}^{(n)}(t)$ are very 
small. Therefore the detector behaves in the same way as if 
the electron stays localized in the left well (Zeno effect). 
A similar behavior would be obtained by introducing the measurement 
collapse. However, the situation
is different when the electron is initially in the statistical mixture, 
or in the linear superposition. For instance, 
solving Eqs.~(\ref{c3a})-(\ref{c3c}) for the initial conditions,
$\sigma_{11}^{(0)}(0)=\sigma_{22}^{(0)}(0)=1/2$ and 
$\sigma_{22}^{(0)}(0)=1/2$ (or $\sigma_{22}^{(0)}(0)=0$),
we obtain that $\sigma_{11}^{(n)}(t)\simeq 
(1/2)\bar\sigma^{(n)}(t)$, and 
$\sigma_{22}^{(0)}(t)\simeq 1/2$ for 
$t{\lower2pt\hbox{$\stackrel{<}{\sim}$}}t_0$. 
It means that at $t\simeq n/D_1$ one can find  $n$ or 
zero electrons in the collector with the same probability $1/2$. 
The first possibility implies that the left dot is occupied, and 
the second one corresponds to the occupied right dot. 
By assuming that one of these possibilities is actually realized,  
one finds the whole system 
starting its evolution from the new initial condition (collapse).
Then the detector current would display quantum jumps, shown in Fig.~2.  
If however, the Schr\"odinger evolution is not interrupted, 
the detector would show the current either $I_1$ or $0$ with 
the probability $1/2$. This means the ``telegraphic'' noise for 
$t{\lower2pt\hbox{$\stackrel{<}{\sim}$}}t_0$. Such a behaviour is 
different from the recent result\cite{leo}, based 
on continuous Schr\"odinger evolution. 
For $t\gg t_0$, however, Eqs.~(\ref{c3a}-\ref{c3c}) 
give the average value $I=I_1/2$ for the detector current, Fig.~2.
\vskip0.2cm
\begin{minipage}{13cm}
\begin{center}
\leavevmode
\epsfxsize=9cm
\epsffile{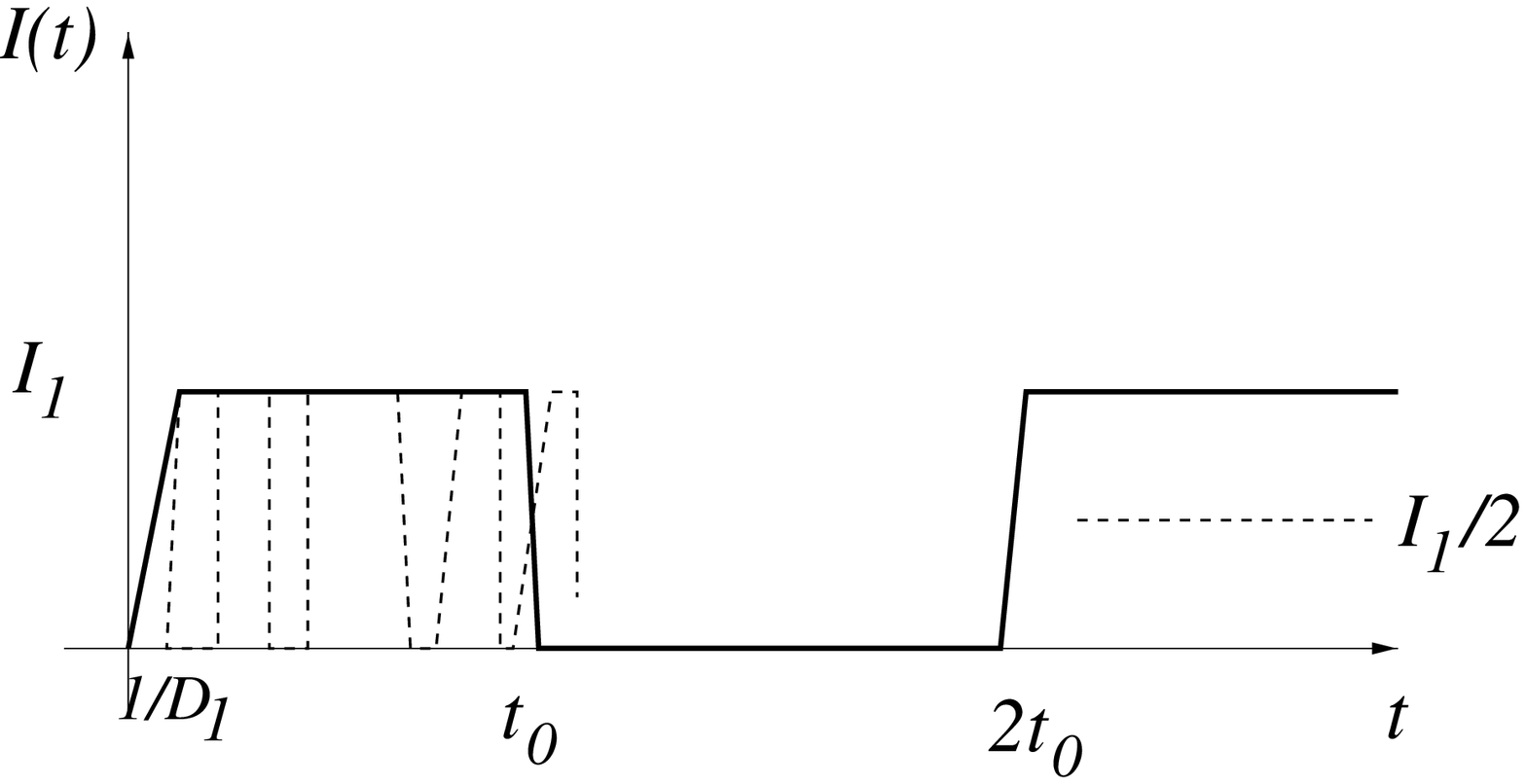}
\end{center}
{\begin{small}
Fig.~2. Detector current as a function of time by 
assuming the collapse (the solid line) and without the collapse 
(the dashed line). The electron is initially in the statistical 
mixture or in the linear superposition.
\end{small}} 
\end{minipage} \\ \\ 

Special thanks to E. Buks for attracting my attention 
to the problem of detector current and numerous very fruitful 
discussions. I am also grateful to Yu. Nazarov for useful discussions. 
The part of this work has been done during my
visit to Delft University of Technology, and I acknowledge  financial
support from the "Nederlandse Organisatie voor Wetenschappelijk Onderzoek"
(NWO), that made this visit possible.

\end{document}